\documentclass[11pt, onecolumn]{article}
\usepackage{hyperref}
\usepackage[cmex10]{amsmath}
\usepackage{amsthm,amssymb,graphicx}
\newtheorem{lemma}{Lemma}
\newtheorem{proposition}{Proposition}
\newtheorem{corollary}{Corollary}
\newtheorem{theorem}{Theorem}
\newtheorem{definition}{Definition}
\newtheorem{observation}{Observation}

\newtheorem{fact}{Fact}
\usepackage{url}

\usepackage{enumitem}
\usepackage[margin=0.9in]{geometry}

\newcommand{\eps}{\epsilon}
\newcommand{\Z}{\mathbb{Z}}
\newcommand{\E}{\textbf{E}}
\newcommand{\Var}{\textbf{Var}}
\newcommand{\A}{\mathcal{A}}
\newcommand{\B}{\mathcal{B}}

\author{ Gregory Valiant \\gregory.valiant@gmail.com
  \and Paul Valiant \\pvaliant@gmail.com}

\date{}

\begin{document}

\title{Information Theoretically Secure Databases}

\maketitle
\thispagestyle{empty}

\vspace{-.15in}\begin{abstract}
We introduce the notion of a database system that is information theoretically \emph{secure in between accesses}---a database system with the properties that 1) users can efficiently access their data, and 2) while a user is not accessing their data, the user's information is \emph{information theoretically} secure to malicious agents, provided that certain requirements on the maintenance of the database are realized.  We stress that the security guarantee is information theoretic and everlasting: it relies neither on unproved hardness assumptions, nor on the assumption that the adversary is computationally or storage bounded.

We propose a realization of such a database system and prove that a user's stored information, in between times when it is being legitimately accessed, is information theoretically secure both to adversaries who interact with the database in the prescribed manner, as well as to adversaries who
have installed a virus that has access to the entire database and communicates with the adversary.

The central idea behind our design of an information theoretically secure database system is the construction of a ``re-randomizing database'' that periodically changes the internal representation of the information that is being stored. To ensure security, these remappings of the representation
of the data must be made sufficiently often in comparison to the amount of information that is being communicated from the database between remappings and the amount of local memory in the database that a virus may preserve during the remappings. While this changing representation provably foils the ability of an adversary to glean information, it can be accomplished in a manner transparent to the legitimate users, preserving how database users access their data.

The core of the proof of the security guarantee is the following communication/data tradeoff for the problem of learning sparse parities from uniformly random $n$-bit examples. Fix a set $S \subset \{1,\ldots,n\}$ of size $k$:  given access to examples $x_1,\ldots,x_t$ where $x_i \in \{0,1\}^n$ is chosen uniformly at random, conditioned on the XOR of the components of $x$ indexed by set $S$ equalling 0, any algorithm that learns the set $S$ with probability at least $p$ and extracts at most $r$ bits of information from each example, must see at least $p\cdot \left(\frac{n}{r}\right)^{k/2} c_k$ examples, for $c_k \ge \frac{1}{4}\cdot\sqrt{\frac{(2e)^{k}}{k^{k+3}} }$.  The $r$ bits of information extracted from each example can be an arbitrary (adaptively chosen) function of the entire example, and need not be simply a subset of the bits of the example.

\end{abstract}

\section{Introduction}

\begin{figure}
\begin{center}
\includegraphics[width=.8\linewidth]{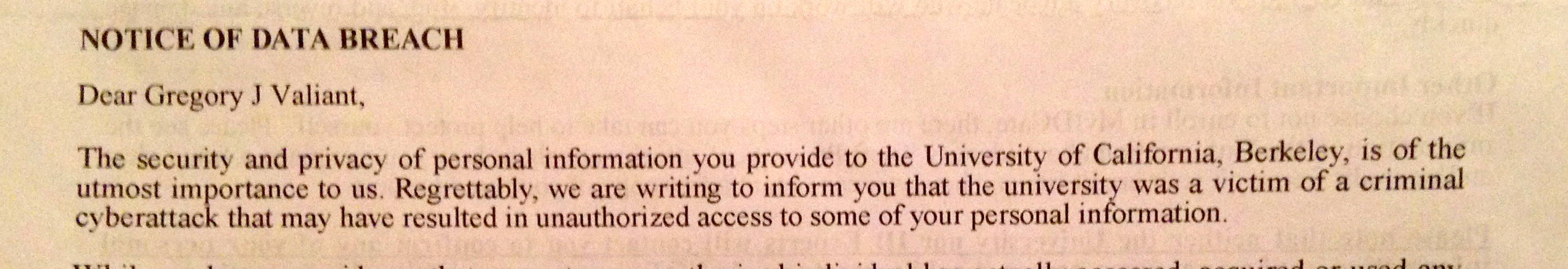}
\includegraphics[width=.8\linewidth]{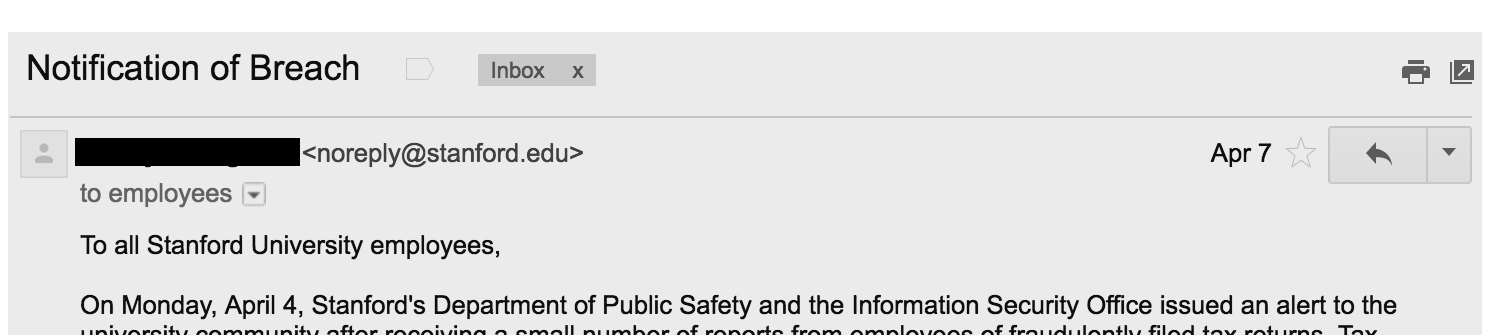}
\caption{Two recent notices that one of the authors received.  In both cases, the personal information was not stolen in transmission---in fact, in the first case, the author has not attempted to access any information from UC Berkeley's computer system during the past 3 years.  A \emph{Secure In Between Accesses} database provides strong defenses against such information theft.\label{fig1}}
\end{center}
\end{figure}

With the increasing aggregation of our sensitive data (medical, financial, etc.) in databases that are accessible over the internet, these databases are increasingly being targeted by malicious agents.  Many of the most worrying and expensive hacks to date have not been due to failures in the transmission of encrypted data, but due to large-scale attacks on the databases themselves.  Stated differently, it is often not the case that a user requests his/her information, and that information is compromised in transmission; instead, in between a user's accesses, an adversary hacks into a database and downloads thousands or millions of sensitive entries.   (See Figure~\ref{fig1} for two  such examples.)  We introduce a notion of database security that stipulates that the data being stored is \emph{secure in between accesses}, which we refer to as ``SIBA security''.

\begin{definition} Given a database system $D$, constructed satisfying some implementation requirements $R$, we say that $D$ is information theoretically secure with unconditional guarantees $G$, if, independent of any assumptions about the computational or storage limitations of outside agents or of the intractability of any functions, some guarantees $G$ for users will be achieved independent of the behavior of outside agents. The guarantees $G$ will be of the form that any user or adversary, even one who has knowledge of some information $I$ in the database, will be unable to learn anything about the information stored in the database beyond $I$, except with a small specified failure probability $\eps$.
\end{definition}

We propose a realization of such an information theoretically secure database, which permits users to access their data efficiently, but which guarantees that any adversary, even one who repeatedly accesses the database, or installs a virus on the database itself, cannot glean any information about the users' data beyond that which it already knows a priori, except with a negligible probability (that we explicitly bound).

The central idea behind our design of an information theoretically SIBA-secure database system is to have the database's representation of the information it stores remap periodically in such a way that 1) honest database users can continue to access their data by XORing together the information at a small unchanging set of addresses, but 2) adversaries are information theoretically
unable to piece together information from the different representations in a sufficiently consistent manner to glean any knowledge. To ensure security, these remappings of the data must occur sufficiently often in comparison to the amount of information that is being communicated from the database.

The implementation requirements, $R$, needed by our realization are threefold. First, the database needs to monitor the total number of bits transmitted from the database to the outside world (whether transmitted in response to queries of legitimate users, arbitrary queries of illegitimate users, or even arbitrary information computed by a virus running on the database) and have the ability to cut the connection to the outside world and re-randomize the database's representation of user data (in a manner transparent to users) every time this transmission limit is exceeded. Second, the database must ensure that at the time of every remapping, for any virus that might be present on the database, the amount of local memory available to it is also bounded by the prescribed bound on the amount of communication from the database.  And third, the database needs to store the addresses and compute the remappings securely.

Our information theoretic security guarantees do not extend to the accessing and transmission of a user's data, or the viewing of the data at the user's end.  At the time of access, a virus in the database may discover the user's addresses in the database and exploit them.   Improving the security of the data during accessing and transmission can be attempted via standard oblivious transfer protocols~\cite{rabin2005exchange,naor2001efficient,lipmaa2005oblivious} and encryption, though it will lack the information theoretic guarantees. For these reasons, we say that our database is ``secure in between accesses''. As long as a user's data is not being accessed, its security is information theoretically guaranteed.

\section{Related Work}

While most security and cryptographic schemes rely on the assumptions that adversaries are computationally bounded and  that certain problems require super-polynomial amounts of computation to solve, there has been significant attention on devising stronger, unconditional security guarantees based on information theory.  This direction, in some sense, began with the proposal of the ``one-time-pad'' in the late 1880's~\cite{miller1882telegraphic} (which was later re-discovered and patented in the 1920's by Vernam~\cite{vernam1926cipher}), and Shannon's ``Communication Theory of Secrecy Systems''~\cite{shannon1949communication} in the 1940's.  More recently, Maurer introduced \emph{bounded-storage cryptography}, that considered cryptographic protocols with information theoretic security, under the assumption that the adversary has limited memory~\cite{maurer1992conditionally} (also see the survey~\cite{maurer1999information}).   There has been a great deal of work in this bounded-storage model, including more recent work on ``everlasting'' encryption in this model, in which information theoretic security persists even if the secret key is divulged at some later point in time (assuming the adversary has bounded storage)~\cite{aumann2002everlasting,ding2002hyper,lu2002hyper}.  Most of the work on bounded memory cryptography assumes that all parties have access to some enormous stream of random bits (for example, random bits that are being continuously broadcast from a satellite), and that the users interact with the stream by selectively storing certain bits.

Our proposal, in which a database monitors the amount of information communicated from the database, and refreshes its representation of the data, introduces a significantly different interaction model from 
previous approaches to information theoretic security and cryptography.    From a technical perspective, both the previous proposals for everlasting encryption protocols as well as our protocol proceed by XORing together the values at a certain secret set of indices---in the case of the everlasting encryption protocols, these index a long stream of random bits, and the XOR is used as a ``pad'' for the bit to be communicated.  Additionally, both our proof of security, as well as that of~\cite{ding2002hyper}, proceed by essentially showing the information theoretic hardness of learning sparse parity functions in a bounded communication setting, discussed below.   What we show is that the combination of re-randomizing the database and controlling the information flow from it offers strong security guarantees of a novel nature.

The core of the proof of the information theoretic security of the database system we propose is a result on the tradeoff between communication and the number of examples necessary to learn a parity function.  There has been significant attention from the learning theory community over the past few decades on understanding such tradeoffs, perhaps beginning with the work of Ben-David and Dichterman~\cite{ben1993learning}.  In the distributed setting, there has been significant recent work analyzing how much data is required to learn certain classes of function in the setting in which data is partitioned across servers, and there are restrictions on the communication between the servers, and/or privacy constraints (e.g.~\cite{BBFM12, DJW13, ZDJW13, GMN14,datproc15,steinhardt2015memory,shamir2014fundamental}).  In many cases, these  results on communication bounded learning immediately yield analogous statements in the memory bounded streaming setting, where the learner has limited memory, and is given access to a stream of examples.

As far as analytic methods, the work most closely related to ours are the works of Ding and Rabin~\cite{ding2002hyper}, Shamir~\cite{shamir2014fundamental}, Steinhardt, Valiant, and Wager~\cite{steinhardt2015memory}, and Raz~\cite{Raz16}.  In~\cite{ding2002hyper}, Ding and Rabin propose a scheme for ``everlasting security'' against storage-bounded adversaries, whose proof of correctness can be essentially described as providing a communication/sample tradeoff for the problem of learning parities of size $k$ from random length $n$ examples.  Their result focuses on the regime in which the communication per example is linear in $n$, specifically $n/6$, and show that the security guarantees will be of the form $exp(k)$.  In~\cite{shamir2014fundamental}, Shamir considers learning with memory and communication constraints, and,  among other results, shows that: for the problem of identifying one significantly biased bit from otherwise uniformly random length $n$ examples,  any learning algorithm with $r$ bits of memory requires $O(n/r)$ examples to correctly identify the biased index.  This corresponds to the problem of learning parities of size $k=1$, from random examples.  In~\cite{steinhardt2015memory}, the authors establish a correspondence between the problems that are learnable via algorithms that communicate/extract few bits of information from each example, and those problems learnable in Kearns' \emph{statistical query} model~\cite{kearns1998efficient}.  Additionally, they show that in the distributed setting in which parties are given examples from a length $n$ instance of parity, either some parties must communicate $\theta(n)$ bits of information about their example, or an exponential number of parties are required to learn the parity set, with high probability.  They also conjectured a stronger result, that for any positive $\eps>0$ and sufficiently large $n$, any algorithm for learning a random parity function over length $n$ examples either requires memory at least $n^2 (\frac{1}{4} - \eps),$ or must see an exponential number of examples to recover the parity set with high probability.  This conjecture was proved by Raz, with the slightly weaker constant of $1/25$ instead of $1/4 - \eps$~\cite{Raz16}.  In that work, Raz also observes that such a result immediately provides an example of a bounded-storage cryptographic scheme where a single bit can be communicated given a length $n$ private key, and time $n$ to encrypt/decrypt, in such a way that it is secure against any adversary with memory less than $n^2/25$.

\section{SIBA--Security via Re-Randomizing Databases}

At a high level, our proposed system works as follows: the database can be regarded as an $n$ bit string $x$, and each user has a specified set of $k$ indices for each bit to be stored.  The user's bit can be accessed by computing the XOR of the database values at the specified $k$ indices.  Periodically, the database will replace $x$ with a uniformly random string, subject to the condition that the XOR of the specified indices is still the desired value.   The security of the system rests on the ability of the database to ensure that a limited number of bits of information have been communicated to the outside world between these ``re-randomizations'' of $x$.   

For clarity, we formally describe the proposed system in the case of a single user, ``Alice'', who wishes to access a single fixed bit of information $b \in \{0,1\}$.

\begin{center}
\hspace{-.1in}\fbox{\parbox{5.4in}{
\begin{sc}An Information Theoretically Secure Database\end{sc}\\
The database will consist of $n$ bits, the parameter $k$ denotes the key size, the parameter $r$ denotes the \emph{database refresh rate}, and Alice wishes to store bit $b \in \{0,1\}$. \medskip

\textbf{Initialization:}
\begin{itemize}
\item Alice and the database agree on a uniformly random secret, $S=\{s_1,\ldots,s_k\} \subset\{1,\ldots,n\}$ with $|S|=k$.
\item The database is initialized to a uniformly random bit string $x \in \{0,1\}^n$ such that $b= \sum_{i \in S} x_{i} \mod{2}$.
\end{itemize}
\textbf{Access:}
\begin{itemize}  \item To access Alice's bit, $b$, she requests the values of $x$ at locations $i \in S$, and computes their XOR.
\end{itemize}
\textbf{Maintenance:}
\begin{itemize}
\item The database counts the total number of bits communicated from the database (in aggregate across all users.)
\item Before this count reaches $r$, the count resets to 0 and the database ``re-randomizes'': it replaces $x$ with a new $x'\in \{0,1\}^n$ chosen uniformly at random conditioned on $b=\sum_{i \in S} x_{i} \mod{2}$.
\end{itemize}
}}\end{center}
\medskip

The crux of the above proposal---that the database ``re-randomizes'' before too much information is transmitted about any given instantiation of the database's memory---can be easily guaranteed in practice by ensuring that two properties hold.   First, that the database is connected to the world via a communication channel with some known bandwidth, and ensuring that the database re-randomizes conservatively, assuming that the channel is permanently transmitting at full capacity.  And second, that there is no large tract of database memory that is skipped in the re-randomizations.   Specifically, to ensure that no adversary extracts too much information about the state of the database at a given time, the database must ensure that any virus that might be present (on the database itself) is restricted to a limited amount of memory between re-randomizations (as such memory could be leveraged at a future time to communicate extra information about the current state of the database).     

\subsection{Time-Varying Data}
The above basic design easily extends to support the setting where Alice's bit changes over time.  Let $b_t$ denote the bit as a function of time, $t$, and assume that time is delimited in discrete intervals.  The above protocol can be trivially adapted to allow Alice access to the time-dependent bit $b_t$, by simply ensuring that the database re-randomizes both at each time increment, and when the communication reaches the database refresh rate parameter, $r$.  With a re-randomization that occurs at time $t$, the database string $x$ is chosen uniformly at random conditioned on the XOR of the indices of $x$ in Alice's secret set equalling $b_t$.   This protocol also extends naturally to the multi-user setting, and the setting where each user stores multiple bits, with security guarantees essentially unchanged, provided the length of the database string is greater than $2k$ times the total number of bits to be stored across all users.  We describe these extensions in Section~\ref{sec:extensions}.

\subsection{Security Guarantees}

The following observation characterizes the security guarantees of the above system in the setting in which an adversary can only interact with the database by requesting the database values at specified indices:

\begin{observation}\label{obs:1}
After at most $t$ ``re-randomizations'' of the database, an adversary that only interacts with the database by requesting the database values at $r$ specified indices per re-randomization can correctly guess Alice's secret subset $S$ with probability at most $t \left(r/n\right)^k$, even if the adversary knows Alice's sequence of bits $b_1,\ldots,b_t$ \emph{a priori}.  Furthermore, an adversary who knows Alice's bit values at times $1,\ldots,i-1,i+1,i+2,\ldots,t$, can distinguish the case that $b_i=0$ from the case that $b_i=1$ with probability at most $t \left(r/n\right)^k.$
\end{observation}

The above observation follows from noting that for any subset of indices $Q \subset \{1,\ldots,n\}$ that does not contain the entire set $S$, the bits of the database at indices in $Q$ will be a uniformly random $|Q|$-length bitstring, and in particular, is independent of Alice's current bit $b_i$ and all of her past and future bits.

The power of the above observation is that even an adversary who knows Alice's data ahead of time cannot leverage this knowledge to any advantage.   In practical terms, for a bank implementing such a system, this would mean that even if an adversary steals a paper copy of Alice's bank account balance and all her historical banking information, the adversary cannot leverage this information to glean any additional information about Alice's account, and, for example, will not be able to detect a change to the account balance, or recover any more historic data than what the adversary already has. Further, this simple setting has the ``everlasting security" property~\cite{aumann2002everlasting} that, if after the database is shut down, the adversary later learns the locations of Alice's secret bit locations, the adversary will not be able to recover any of Alice's secrets (unless, as happens with probability $t(r/n)^k$, during one of the $t$ periods the adversary had previously got lucky and simultaneously observed all $k$ of the bits at Alice's locations).

\medskip

The following theorem, which is our main result, shows that similar information theoretic security persists even in the presence of significantly more pernicious attacks on the database.  Suppose an adversary hacks into the database and installs a virus that allows the adversary to interact with the database in a more general fashion, rather than simply querying the database value at specified indices.  Even if the virus can compute arbitrary functions of the \emph{entire} database and transmit these function evaluations to the adversary, the database will be secure with essentially the same bounds provided at most $r$ bits have been transmitted between ``re-randomizations''.  
This security holds even if the adversary has infinite memory and computational power, and can communicate arbitrary amounts of information to the virus---for example, even in the setting where the adversary is able to upload a new virus with every re-randomization.   Alternatively, instead of assuming a bound of $r$ communication to the outside world, the same results hold when bounding the communication of the virus to its future self: the assumption that the database can ensure that no virus preserves more than $r$ local memory on the database between re-randomizations is practically feasible as the database simply needs to ensure that there is no very-large tract of memory that is left untouched during the ``re-randomizations''.

\begin{theorem}\label{thm:sys}
Given the database system described above with key size $k$ that maintains an $n$-bit string, any algorithm that extracts at most $r$ bits of information about the database between re-randomizations can correctly guess Alice's secret set, $S$, with probability at most ${n\choose k}^{-1}+ t \cdot \left(\frac{r}{n}\right)^{k/2} \cdot 4\sqrt{\frac{k^{k+3}} {(2e)^{k}}}$  after $t$ re-randomizations.   Furthermore, the security of Alice's data is ``everlasting'': suppose Alice stores bits $b_1,\ldots,b_{t-1}$ in the database for the first $t-1$ re-randomizations, and these bits are known to the adversary ahead of time.  If Alice then chooses bit $b_t$ at random from $\{0,1\}$, and the adversary extracts at most $r$ bits of information from the database during each of the first $t$ re-randomizations, then even if the adversary is given Alice's secret set $S$ after the $t+1$st rerandomization, the adversary can correctly guess Alice's bit with probability at most $1/2 + t \cdot \left(\frac{r}{n}\right)^{k/2} \cdot 4\cdot\sqrt{\frac{k^{k+3}} {(2e)^{k}}}$.
\end{theorem}

One reasonable setting of parameters would be to use a database string of size $n=10^{12}$, with rerandomization occurring every $r = 10^8$ bits transmitted, and where each bit stored in the database is represented as the XOR of $k=10$ secret locations. In this case Theorem~\ref{thm:sys} guarantees information theoretic security except with probability $<3\cdot 10^{-17}$, per database rerandomization, for each bit stored in the database.

The above theorem can also be viewed as a communication/data tradeoff for the problem of learning $k$-sparse parities over $n$-bit examples:

\begin{corollary}\label{cor:sparsep}
Choose a uniformly random set $S=\{s_1,\ldots,s_k\} \subset \{1,\ldots,n\}$ of size $k$:  given access to a stream of examples $x_1,\ldots,x_t$ where each $x_i \in \{0,1\}^n$ is chosen uniformly at random, conditioned on the XOR of the components of $x_i$ with indices in $S$ being $b_i$, any algorithm, given $b_1,\ldots,b_t$, that must (even adaptively) compress each example to at most $r$ bits before seeing the next example can correctly guess the set $S$ with probability at most ${n\choose k}^{-1}+t \cdot \left(\frac{r}{n}\right)^{k/2} \cdot 4\sqrt{\frac{k^{k+3}} {(2e)^{k}}}$.
\end{corollary}

This result can be viewed as mapping the intermediate regime between the communication/data tradeoffs given by Ohad Shamir~\cite{shamir2014fundamental} for the case $k=1$ (the ``hide-and-seek'' problem of detecting a biased index from otherwise uniformly random length $n$ sequences), and the results in Steinhardt, Valiant, and Wager~\cite{steinhardt2015memory} for the case $k=\theta(n)$, which shows that any algorithm that extracts less than $n-c$ bits of information from each example, must see at least $2^{\theta(c)}$ examples.  

The proof of Theorem~\ref{thm:sys} is given in Section~\ref{proof}.

\section{Extension to Multiple Users and Multiple Bits}\label{sec:extensions}

We present the straightforward extension of our secure database to the setting where there are multiple users, each storing multiple bits.  Each bit to be stored will have its own associated secret set of $k$ indices, disjoint from the sets corresponding to all other bits (whether from the same user or a different user). To construct a length-$N$ string that stores $s$ bits collectively, across various users: for each secret $b_i$, to be stored in locations $h_{i,1},\ldots,h_{i,k}$, the database independently chooses a random set of $k$ bits of parity $b_i$ and assigns them to the locations $h_{i,1},\ldots,h_{i,k}$ in the string; the remaining locations in the string are chosen to be independent coin flips. The security guarantees on a length-$N$ database storing $s$ bits collectively across all users result from the following observation: for each secret bit of a user, even assuming the adversary has complete information about all aspects of all $s-1$ remaining bits (from this user and the other users), then the database setup for the remaining secret bit is effectively identical to the standard setup of Theorem~\ref{thm:sys} for a database with string length $n=N-(s-1)k$.

 Thus, provided the size of the database representation $N$, is at least "n-k" bits larger than $k$ times the number of bits being stored, the security guarantees of Theorem~\ref{thm:sys} persist. Even in the event that an adversary has prior knowledge of some of the secret sets and bits, with all but negligible probability, the adversary cannot leverage this knowledge to gain any additional knowledge:

\begin{corollary}
Consider a database system as described above that maintains a length $N$ string, has key size $k$ (per bit stored), and stores a total of $s$ bits $b_1,\ldots,b_s$, with the $i$th bit corresponding to the XOR of $k$ indices $h_i \subset \{1,\ldots,n\}$, where the sets $h_i$ are disjoint.  The security guarantees of Theorem~\ref{thm:sys} hold for $n=N-(s-1)k$, for each given bit $b_i$ and secret $S_i$, even if the adversary knows partial or complete information about the remaining $s-1$ bits $b_j$ and secret key sets $h_j$.
\end{corollary}
\begin{proof}
  We consider the case when the adversary has \emph{complete} information about the remaining $s-1$ bits and secret key sets, as such an adversary can accomplish at least as much as one with only partial information. Consider ignoring all the bits in the $(s-1)k$ known secret key locations $h_{2,1},\ldots,h_{s,k}$ from the database string of length $N$, the (joint) distribution of what remains is identical to the construction of a single bit in a database of size $n=N-(s-1)k$, since each of the secret key locations is chosen independently and disjointly, and for each $k$-tuple of secret locations the $k$ bits at these locations are chosen independently of the rest of the database, and each bit not at a secret location is chosen by an independent coin flip.
   \end{proof}
   The above proof can alternatively be viewed as follows:  given a database of size $n$ securely storing a single bit, the adversary could easily simulate having access to a database of size $N = n+(s-1)k$ securely storing $s$ bits, where the adversary knows everything about the $s-1$ simulated secrets (because the adversary simulated adding them to the database).  If there were any way to extract information about one of the bits in the $s$-bit setting by leveraging partial or complete information about the other $s-1$ bits, then an adversary could simulate this attack in the single-bit setting, contradicting Theorem~\ref{thm:sys}.

\subsection{Decreasing the Key Size}\label{sec:shorter-key}
In this multiple-bit setting, as described above, a user will store $sk$ secret indices for every $s$ bits of information that she wishes to store securely.  There are many natural approaches to improving this scaling of parameters, including analogs of the pseudorandom generators induced by random walks that were used in a related setting to substantially decrease the key size for a user, so as to be sublinear in the number of bits she wants to store~\cite{lu2002hyper}. 

\section{Secure Communication via Re-Randomizing Databases}\label{sec:multiuse}
A variant of the concept of a re-randomizing database can also be leveraged to provide a secure channel between pairs of people that replicates the guarantees of a one-time pad, but allows the pair to securely and efficiently generate new one-time pads.  In this setting, the ``database'' is just a source of randomness, and does not need any secure storage.  One can regard the following protocol as a variant of the setting considered by the work on ``everlasting'' encryption~\cite{aumann2002everlasting}.  That work assumes that there is a source broadcasting  random bits (e.g. a satellite broadcasting a high-rate stream of random bits) and storage-bounded adversaries; here instead, in our setting the random bits are available on a database, and the database re-randomizes the bits after a fixed amount of information has been transmitted.  We describe the protocol below in the setting where  Alice wishes to communicate one bit of information, $b_t$ at each timestep $t$, to Bob.

\begin{center}
\hspace{-.1in}\fbox{\parbox{5.4in}{
\begin{sc}The Refreshable One-Time Pad\end{sc}\\
The ``database'' will consist of $n$ bits, the parameter $r$ denotes the \emph{database refresh rate},  the parameter $k$ denotes the key size. \medskip

\textbf{Alice and Bob Initialization:}
\begin{itemize}
\item Alice and Bob securely exchange a key set $S$ consisting of a uniformly random set of $k$ elements of $\{1,\ldots,n\}$.
\end{itemize}

\textbf{Database Maintenance:}
\begin{itemize}
\item The database initializes to a uniformly random length $n$ vector $x \in \{0,1\}^n$.
\item The database counts the total number of bits communicated from the database (in aggregate across all users.)
\item Before this count reaches $r$, the count resets to 0 and the database ``re-randomizes'': it replaces $x$ with a new $x'\in \{0,1\}^n$ chosen uniformly at random.
\end{itemize}

\textbf{Alice and Bob Communication:}
\begin{itemize}  \item To communicate bit $b_t$ to Bob, Alice queries the database at locations corresponding to indices in $S$.  She then sends Bob  (over an insecure channel) the XOR of these bits with her message, $b_t$.   Bob then queries the database at the indices specified by $S$, and computes $b_t$ by XORing the message with the XOR of the retrieved bits.  Alice and Bob may send multiple messages between re-randomizations of the database by sharing multiple secret sets, $S_1,S_2,$ etc. and ensuring that they do not use the same secret set $S$ multiple times between database re-randomizations.
\end{itemize}
}}\end{center}
\medskip

There are two obvious security weaknesses of the above communication protocol.  The first is that the database must be trusted to re-randomize.  The second weakness applies even to an honest-but-curious database: the communication scheme, as stated above, discloses the secret set, $S$, to the database, allowing the database to decrypt Alice's message.   To partially address this problem, both Alice and Bob could elicit the values at locations indexed by set $S$ via a ``1--of--n'' \emph{oblivious transfer} protocol (see e.g.~\cite{rabin2005exchange,naor2001efficient,lipmaa2005oblivious}).  Such a protocol guarantees that the database does not learn the indices queried (or resulting values), and that Alice and Bob do not learn anything about the database other than their desired values.  These guarantees are not information theoretic, and instead rely on cryptographic assumptions (and the assumption that the parties are computationally bounded).  Nevertheless, our ``everlasting'' information theoretic guarantees at least yield the fact that, unless the adversary successfully decrypts the oblivious transfer \emph{before} the database re-randomizes, Alice's message will always be secure.  In this sense, the above protocol implemented via oblivious transfers at least has the property that one only needs to ensure that the oblivious transfer is secure \emph{against the current state-of-the-art}.  One does not need to worry about the development of future technology (quantum computers, new attacks, etc.)---by the time such attacks are developed, the ``database'' will have re-randomized, and the information necessary to decrypt Alice's message will have been lost.

\section{Extensions to Larger Fields}
In addition to the generalization discussed in Section~\ref{sec:shorter-key} for shortening the key size, there are several other basic generalizations of our specific realization of an information theoretically secure database.  One such generalization is to have the internal representation correspond to elements of an arbitrary group $G$.  (For example, $G= \Z_m$, for an integer $m > 2$, as opposed to the $\Z_2$ version of the database system described above.)   In these larger groups, the database construction and maintenance would be analogous to the $\Z_2$ setting: a user wishes to access some value $b \in G$, and the database would maintain a string $x \in G^n$.  The user would have a set of $k$ indices $S \subset \{1,\ldots,n\}$, and the database would ensure at each re-randomization that $x$ is drawn uniformly at random from $G^n$, subject to $\sum_{i \in S} x_i = b$ (where the arithmetic is with respect to the group operation).  Such a system would have analogous security properties, and would benefit from storing more information for a given size of the secret set, $S$.

\section{Proof of Theorem~\ref{thm:sys}}~\label{proof}
We begin with a high level overview of the proof of Theorem~\ref{thm:sys}.  Because our proof relies on properties of polynomials of random $\pm 1$ variables, which lets one express the parity of $k$ bits as a degree $k$ monomial, for the entirety of this section we will refer to bits as being $\pm 1$ valued rather than $0/1$ valued (as was done in the rest of the paper).  Given an $n$-bit database $dat$ from which an adversary runs an arbitrary computation returning an $r$-bit output $OUT_{dat}$, the challenge is to show that $OUT_{dat}$ gives essentially no information about either of the two aspects of the database we wish to keep secret: the user's secret key, specified by $k$ locations $h_1,\ldots,h_k$; and the user's secret itself, which is stored as the XOR of these $k$ locations in the database. Consider the portion of the hypercube of possible databases $dat\leftarrow\{-1,1\}^n$ that induces a particular output $OUT$---because there are only $2^r$ possible $r$-bit outputs, a typical output $OUT$ must be induced by a relatively large, $2^{-r}$ fraction of the hypercube. The main technical step is arguing why, for any large subset of the hypercube, most $k$-bit parities will return almost exactly as many 1's as $-1$'s on this set. In other words, the only subsets of the hypercube that are biased for many parities are those subsets that are very small. A singleton set is biased for \emph{all} parities but consists of a $2^{-n}$ fraction of the cube; the set of points whose first $k$ coordinates are 0 is a fairly large fraction of the hypercube, but is only strongly biased for the $k$-way parity consisting of exactly the first $k$ bits; in general, large subsets of the hypercube are very close to unbiased, on a typical parity. We analyze this situation in Proposition~\ref{thm:blah}.

Given this tool, the single time-step ($t=1$) special instance of Theorem~\ref{thm:sys} follows by the straightforward argument that, even given $r$ bits of output from the database, the joint conditional distribution of the $k$ secret locations, and XOR of the values in these locations is very close to uniform. Following the intuition of~\cite{lu2002hyper}, this implies both that 1) If the adversary has $r$ bits of output and somehow knows the user's secret data, then the secret key is still close to independent of this information, and thus remains secure; 2) If in addition to the $r$ bits of output, the adversary somehow, after the database has closed down, learns the user's secret key, then the user's secret data remains independent of this information, and thus has ``everlasting'' security--namely, with high probability it is information theoretically impossible to learn Alice's bit.  To obtain the proof of Theorem~\ref{thm:sys} for general $t$, we proceed by induction on $t$, leveraging the single time-step special case.   The details of this proof overview are given below.

\subsection{Large Sets Have Few Biases}

In this section we show that for any sufficiently large section of the hypercube, relatively few sized $k$ parities may have a significant bias.  We begin by formalizing the notion of ``bias'' in a slightly more general setting.

\begin{definition}
Given a function $f:\{-1,1\}^n\rightarrow [0,2^{-n}]$ and a $k$-tuple of indices $h\subset \{1,\ldots,n\}$, the \emph{bias} of $f$ with respect to $h$ is the average value of the degree $k$ monomial induced by $h$ on the conditional distribution induced by $f$.  Formally, $$bias(h,f) = \frac{1}{|f|}\sum_{x \in \{-1,1\}^n} f(x) \prod_{i \in h} x_i,$$ where $|f| = \sum_{x \in \{-1,1\}^n} f(x).$
\end{definition}

The following proposition shows that no function $f:\{-1,1\}^n\rightarrow [0,2^{-n}]$ can have a significant bias with respect to too many $k$-tuples.

\begin{proposition}\label{thm:blah}
Let $S$ denote the set of all $k$-tuples of indices in $\{1,\ldots,n\}$ (hence $|S|={n \choose k}$).  For an even integer $k$, given a function $f:\{-1,1\}^n\rightarrow [0,2^{-n}]$,  the sum over all $h \in S$ of the square of the bias of $f$ with respect to $h$ is bounded as: $$\sum_{h \in S} bias(h,f)^2 \le\frac{ 4 k^{k+3} (2-\log |f|)^k}{(2e)^k},$$ where $|f| = \sum_{x \in \{-1,1\}^n} f(x).$
\end{proposition}

Our proof will leverage the following hypercontractivity concentration inequality from~\cite{schudy2012concentration} (see~\cite{ODonnell-Notes,GaussianHilbertSpaces} for details):
\begin{theorem}[Thm 1.10 from~\cite{schudy2012concentration}]\label{tailb}
For any degree $k$ polynomial $P(x)=P(x_1,\ldots,x_n),$ where the $x_i$ are independently chosen to be $\pm 1$, $$\Pr_{x \in \{-1,1\}^n}\left[\left|P(x) - \E[P(x)]\right| \ge \lambda \right] \le e^2 \cdot e^{-\left(\frac{\lambda^2}{e^2 \Var[P(x)]}\right)^{1/k}},$$
\end{theorem}

Additionally, we will leverage the following standard fact about the upper incomplete gamma function:
\begin{fact}\label{gamma}
Letting $\Gamma(s,\alpha) = \int_{t=\alpha}^{\infty}t^{s-1}e^{-t}dt$ denote the upper incomplete gamma function, for any positive integer $s$, $$\Gamma(s,\alpha) = (s-1)! e^{-\alpha} \sum_{i=0}^{s-1} \frac{\alpha^i}{i!}.$$
\end{fact}

\begin{proof}[Proof of Proposition~\ref{thm:blah}]
Define $P(x) = \sum_{h \in S} bias(h,f) \prod_{i \in h} x_i$ to be the degree $k$ polynomial with $|S|={n \choose k}$ monomials, with coefficients equal to the biases of the corresponding monomials/sets.   Let $s =  \sum_{h \in S} bias(h,f)^2$ denote the quantity we are trying to bound, and note that $$s|f| = \sum_{x\in \{-1,1\}^n}f(x) \cdot P(x).$$

To bound this sum, given the polynomial $P$, consider the function $f^*: \{-1,1\}^n \rightarrow [0,2^{-n}]$ with $|f^*| = |f|$ that maximizes the above quantity.  Such an $f^*$ can be constructed by simply sorting the points of the hypercube $x_1,x_2,\ldots,x_{2^n}$ s.t. $P(x_i) \ge P(x_{i+1}),$ and then setting $\frac{1}{2^{n}} = f^*(x_1)=f^*(x_2) = \ldots = f^*(x_{j})$ for $j = |f| / 2^n$, and $f(x_i)=0$ for all $i > |f|/2^n$.  (For simplicity, assume $|f|$ is a multiple of $1/2^n$; if this is not the case, then we set $j=\lfloor |f| / 2^n\rfloor$ and $f^*(x_{j+1}) = |f| - j/2^n$ and the argument proceeds analogously.)  We now bound
$$\sum_{x\in \{-1,1\}^n}  f(x)P(x) \le \sum_{x\in \{-1,1\}^n}  f^*(x)P(x) = \sum_{\lambda: \exists j \le |f| / 2^n \text{ with }  P(x_j) = \lambda} \lambda \cdot \Pr_{x  \leftarrow \{-1,1\}^n} [ P(x) = \lambda],$$ where the probability is with respect to the uniform distribution over $x\in \{-1,1\}^n$.   Given any differentiable function $g(\lambda)$ that satisfies $ \Pr_{x  \leftarrow \{-1,1\}^n} [ P(x) \ge \lambda] \le g(\lambda),$ we have
\begin{eqnarray}\sum_{\lambda:\exists j \le |f| / 2^n \text{ with }  P(x_j) = \lambda} \lambda \cdot \Pr_{x  \leftarrow \{-1,1\}^n} [ P(x) = \lambda] \le \int_{\lambda_0}^{\infty} \lambda \cdot \left|\frac{d}{d\lambda}g(\lambda)\right| d\lambda, \label{eq1}\end{eqnarray}
 where $\lambda_0$ is chosen to be the largest value that $\lambda$ can take, such that $g(\lambda_0) \ge  |f|.$  By Theorem~\ref{tailb}, we may take $$g(\lambda) =  e^2 \cdot e^{-\left(\frac{\lambda^2}{e^2 \Var[P(x)]}  \right)^{1/k}} = e^2 \cdot e^{-\left(\frac{\lambda^2}{e^2\cdot s}  \right)^{1/k}}.$$    Hence taking $\lambda_0$ so as to satisfy $|f| = e^2 \cdot e^{-\left(\frac{\lambda_0^2}{e^2 \cdot s}  \right)^{1/k}},$  yields $$\lambda_0 = (2-\log |f| )^{k/2} \left(e^2 \cdot s\right)^{1/2}.$$   Plugging this into Equation~\ref{eq1} and noting that $\lambda | \frac{d}{d \lambda}g(\lambda)| = \frac{2}{k} \left(\frac{\lambda^2}{e^2 \cdot s}\right)^{1/k}e^{-\left(\frac{\lambda^2}{e^2  \cdot s}\right)^{1/k}}$, we get the following:
\begin{eqnarray*}
\sum_x f(x)P(x) &  \le & \frac{2 e^2}{k}\int_{\lambda_0}^{\infty} \left(\frac{\lambda^2}{e^2 \cdot s}\right)^{1/k}e^{-\left(\frac{\lambda^2}{e^2 \cdot s}\right)^{1/k}} d\lambda \qquad \text{making the substitution } u = \left( \frac{\lambda^2}{e^2  \cdot s}\right)^{1/k}\\
& = & e^2 \left(e^2  \cdot s\right)^{1/2}\int_{ u_0}^{\infty}u^{k/2} e^{-u} du  \qquad \qquad \text{for } u_0 = \left( \frac{\lambda_0^2}{e^2  \cdot s}\right)^{1/k} = 2-\log|f|\\
& = & e^3 \sqrt{s} \Gamma(k/2+1,u_0) \\
& = &  e^3 \sqrt{s}  (k/2)! e^{-u_0} \sum_{i=0}^{k/2} \frac{u_0^i}{i!} \\
& \le & e \sqrt{s}  |f|  (k/2)!(k/4)(2-\log |f|)^{k/2}.
\end{eqnarray*}

The above establishes that $s |f| \le e \sqrt{s}  |f|  (k/2)!(k/4)(2-\log |f|)^{k/2} \le 2 \sqrt{s}  |f| (2e)^{-k/2} k^{k/2+3/2}(2-\log|f|)^{k/2},$ which implies that $s \le 4 (2e)^{-k} k^{k+3} (2-\log |f|)^k,$ as desired.
\end{proof}

\subsection{Completing the Proof}

Equipped with Proposition~\ref{thm:blah}, we now analyze the overall behavior of our secure database. We begin by proving that the security holds for a single re-randomization of the database, and then leverage that result via a basic induction argument to show that the security guarantees degrade linearly with the number of re-randomizations.  The argument of this section closely follow the proof approach of~\cite{lu2002hyper}.

 We begin by considering an adversary that, given the $n$ bits contained in the database, conducts an arbitrary computation to produce an output $OUT$ that is $r$ bits long, and show that, over the random choice of the $k$ locations $h_1,\ldots,h_k \in [n]$ and the random choice of the database $dat\in\{-1,1\}^n$, even given $OUT$, the joint distribution of 1) the $k$ locations $h_1,\ldots,h_k$ and 2) the parity of these $k$ locations, is very close to being jointly uniform and independent.

Using the notation $\langle OUT_{dat}, h_1\ldots h_k, dat_{h_1}\oplus\cdots\oplus dat_{h_k}\rangle$ to represent the joint distribution of these three random variables, and letting $U^h$ and $U^{\pm 1}$ denote the uniform distribution over the set $S=\{h_1,\ldots,h_k\} \subset [n]^k$ and the uniform distribution on $\pm 1$, respectively, we have the following immediate corollary of Proposition~\ref{thm:blah}, which shows the joint distributions $\langle OUT_{dat}, h_1\ldots h_k, dat_{h_1}\oplus\cdots\oplus dat_{h_k}\rangle$ and $\langle OUT_{dat}, U^h,U^{\pm 1}\rangle$ are exponentially close. This implies that, even with the hints provided by $r$ bits of output $OUT$, 1) knowing the user's secret data $dat_{h_1}\oplus\cdots\oplus dat_{h_k}$ gives exponentially little information about the secret key $h_1\ldots h_k$ implying that the key can be securely reused an exponential number of times; and 2) if after the database closes, the secret key $h_1\ldots h_k$ is revealed, everlasting security still holds and the adversary has exponentially little information about the user's secret data $dat_{h_1}\oplus\cdots\oplus dat_{h_k}$, which implies the main results of this paper.

\begin{lemma}\label{lemma:1dat}
  The statistical distance between the distributions $\langle OUT_{dat}, h_1\ldots h_k, dat_{h_1}\oplus\cdots\oplus dat_{h_k}\rangle$ and $\langle OUT_{dat}, U^h,U^{\pm 1}\rangle$ induced by randomly drawing $dat\leftarrow \{-1,1\}^n$ is at most $\left(\frac{r}{n}\right)^{k/2} \cdot \frac{2 k^{k/2+3/2}} {(2e)^{k/2}}$.
\end{lemma}
\begin{proof}
  For a fixed $r$-bit string $OUT_{dat}$, consider the function $f_{OUT}:\{-1,1\}^n \rightarrow [0,2^{-n}]$ that on each string $x\in\{-1,1\}^n$ takes value equal to the joint probability that $x$ is the chosen $n$-bit string \emph{and} that the $r$ bit output string equals $OUT_{dat}$. Proposition~\ref{thm:blah} yields that, $$\sum_{h \in S} bias(h,f_{OUT})^2 \le c_k (2-\log |f_{OUT}|)^k,$$ where $c_k = \frac{4 k^{k+3}} {(2e)^k}$, and $|f_{OUT}| = \sum_{x \in \{-1,1\}^n} f_{OUT}(x).$

  Combining this result with the Cauchy-Schwarz inequality relating the sum of the elements of a vector to the sum of the squares of its elements, we have $$\sum_{h \in S} bias(h,f_{OUT}) \le \sqrt { {n\choose k}\cdot c_k (2-\log |f_{OUT}|)^k}.$$

  We observe that $|f_{OUT}|$, by definition, equals the probability that the particular value of $OUT$ is chosen from among all $r$-bit strings; further, for this fixed $OUT$, the statistical distance between the joint distribution $\langle h_1\ldots h_k, dat_{h_1}\oplus\cdots\oplus dat_{h_k}\rangle$ and the corresponding uniform distribution $\langle U^h,U^{\pm 1}\rangle$ equals ${n\choose k}^{-1}\cdot\sum_{h\in S}bias(h,f_{OUT})$.

  Thus the desired statistical distance between the distributions $\langle OUT, h_1\ldots h_k, dat_{h_1}\oplus\cdots\oplus dat_{h_k}\rangle$ and $\langle OUT, U^h,U^{\pm 1}\rangle$ is bounded by \[\mathop{\mathbb{E}}_{OUT} \left[\sqrt { {n\choose k}^{-1}\cdot c_k  (2-\log |f_{OUT}|)^k}\right]=\sum_{OUT} |f_{OUT}|\cdot\sqrt { {n\choose k}^{-1}\cdot c_k  (2-\log |f_{OUT}|)^k},\] subject to the constraint that $\sum_{OUT} |f_{OUT}|=1$. Since $x(2-\log x)^{k/2}$ is a concave function of $x$ for $x\in[0,1]$, the statistical distance is thus maximized when for each of the $2^r$ possible outputs $OUT$, the probabilities are all equal: $|f_{OUT}|=2^{-r}$. Plugging this in to the above equation gives the desired bound on the statistical distance: \begin{align*}{\big |}\langle OUT_{dat}, h_1\ldots h_k, dat_{h_1}\oplus\cdots\oplus dat_{h_k}\rangle & - \langle OUT_{dat}, U^h,U^{\pm 1}\rangle{\big |} \\ &\leq \sqrt { {n\choose k}^{-1}\cdot c_k (2-\log 2^{-r})^k}=\left(\frac{r}{n}\right)^{k/2}\cdot  \frac{2 k^{k/2+3/2}} {(2e)^{k/2}}.\end{align*}
\end{proof}

We now complete the proof of our main security guarantee, Theorem~\ref{thm:sys}, which we restate below in the above terminology. We use the notation $[t]$ for an integer $t$ to denote the set $\{1,\ldots,t\}$. We show that an adversary repeatedly hijacking the database essentially learns nothing beyond what the adversary could have learned by staying home and simulating the whole process. This guarantee holds even if the adversary finds out about all of Alice's previously stored bits, or, more generally, receives arbitrary ``hints" from an outside source about Alice's past, present, and future bits. We proceed to show the information theoretic security of our database scheme by showing that for any adversary extracting information from the database, there is an analogous \emph{simulator} that the adversary could run without any access to the database, whose results are identical with all but negligible probability.  Such simulator constructions were originally developed and employed in the context of semantic security~\cite{GoldwasserMicali}.\medskip

\noindent \textbf{Theorem~\ref{thm:sys}.} \emph{For any adversary, there is an efficient simulator $S$ such that for any sequence of bits $b_i$ to be stored at a succession of rerandomization times in the database, and any sequence of (possibly probabilistic) ``hints" $H_i$ that the adversary receives about the (previous, current, or future) bits in the sequence, then, averaged over the all $n\choose k$ secret $k$-tuples of  locations $h_{[k]}$, the statistical distance between the distribution of the view of the adversary after running on the database for $t$ rounds, receiving hint $H_i$ after each round $i$ versus the view of the simulator who is given hints $H_{[t]}$ but {\bf never} interacts with the database, is  less than $2t \cdot \eps_{r,n,k}$, where $\eps_{r,n,k} = \left(\frac{r}{n}\right)^{k/2} \cdot \sqrt{\frac{4 k^{k+3}} {(2e)^{k}}}$ is the bound given in Lemma~\ref{lemma:1dat} for a single re-randomization.}


  \medskip

  This theorem has the following immediate interpretations:
  \begin{enumerate}
  \item If at the end of $t$ database rerandomizations an adversary is told Alice's bits $b_1,\ldots,b_{t}$, then it still cannot guess Alice's secret indices correctly with probability any better than $2t\eps_{r,n,k}$ more than random guessing.
\item  If the database represents a uniformly random bit $b_t \in \{-1,1\}$ during the $t$th re-randomization, then even if an adversary is told (at the very beginning) the $t-1$ bits, $b_1,\ldots, b_{t-1},$ that Alice is storing during the first $t-1$ database rerandomizations, and even if, subsequent to the $t+1$st rerandomization, the adversary is told Alice's secret set of indices, then the adversary can guess $b_t$ correctly with probability at most $2t\eps_{r,n,k}$ better than random guessing. This is the ``everlasting security" property.
  \end{enumerate}


\begin{proof}
We prove the theorem by induction on the number of rerandomizations $t$, where the $t=0$ case corresponds to 0 rounds of the database, where the theorem trivially holds since neither the real nor simulated adversary has any information.

Assume, by the induction hypothesis, that there is an efficient simulator $S$ that on input $H_{[t-1]}$ can probabilistically construct a sequence of outputs $OUT'_{[t-1]}$ that is (averaged over all $n
 \choose k$ choices of secret bit locations $h_{[k]}$) within statistical distance $2(t-1)\eps_{r,n,k}$ of the distribution of outputs $OUT_{[t-1]}$ produced by an adversary running for $t$ rerandomizations on the actual database that encodes Alice's secret bits $b_{[t]}$ in secret locations $h_{[k]}$. We couple the random variables $OUT_{[t-1]}$ and $OUT'_{[t-1]}$ together so that they differ with probability $\leq 2(t-1)\eps_{r,n,k}$.

 When the adversary is running on the database during the $t$th rerandomization, it calculates the $t$th output via some function $OUT_t = f(dat_t,OUT_{[t-1]},H_t)$, in terms of the current database (which was randomly drawn so as to encode Alice's bit $b_t$ as the XOR of locations $h_{[k]}$), the previous outputs, and whatever ``hint" $H_t$ it receives about Alice's bits. We change the distribution of $OUT_t$ with probability $\leq 2(t-1)\eps_{r,n,k}$ if we modify it to a ``primed" version $OUT'_t = f(dat_t,OUT'_{[t-1]},H_t)$.

 Since $OUT'_{[t-1]}$ is constructed by the simulator, it is independent of the locations $h_{[k]}$, though possibly dependent on Alice's current secret bit $b_t$ (through hints the adversary received). Thus the output $OUT'_t = f(dat_t,OUT'_{[t-1]},H_t)$ is a function of $dat_t$, independent of the locations $h_{[k]}$, possibly dependent on bit $b_t$ (and also possibly dependent on previous bits $b_1,\ldots,b_{t-1}$ and future bits $b_{t+1},\ldots$, though these do not matter here); we thus denote $OUT'_t = f_{b_t}(dat_t)$, where the function $f_{b_t}$ is possibly stochastic. We thus apply Lemma~\ref{lemma:1dat} to both $f_{b_t=-1}$ and $f_{b_t=1}$: we interpret here interpret Lemma~\ref{lemma:1dat} as saying that for any function $f$ that outputs $r$ bits, the average over all choices of secret locations $h_{[k]}$ and both choices of the bit $b_t$ of the statistical distance between the output of $f$ applied to a database generated from $h_{[k]}$ and $b_t$ versus the output of $f$ when applied to a uniformly random string $dat\leftarrow\{-1,1\}^n$ is at most $\eps_{r,n,k}$.

 Since this bound of $\eps_{r,n,k}$ is averaged over both choices of the bit $b_t$, we bound the statistical distance for \emph{either} choice by twice this, $2\eps_{r,n,k}$. Thus, for both $b_t=-1$ and $b_t=1$ we have that, averaged over all choices of secret locations $h_{[k]}$, the statistical distance between $f_{b_t}$ when evaluated on a database generated from the secrets $h_{[k]}$ and $b_t$ versus when $f_{b_t}$ is evaluated on a uniformly random string $dat\leftarrow\{-1,1\}^n$ is at most $2\eps_{r,n,k}$

 Thus, our simulator, after having already simulated $OUT'_{[t-1]}$ (by the induction hypothesis), next simply draws a random string $dat_t\leftarrow\{-1,1\}^n$ and lets $OUT'_t=f(dat_t,OUT'_{[t-1]},H_t)$.
 The coupling argument shows that the first $t-1$ outputs are accurately simulated except with probability $\leq 2(t-1)\eps_{r,n,k}$, and provided the first $t-1$ outputs are accurately simulated, the previous paragraph shows that the $t$th output has the desired distribution, up to statistical distance error $\leq 2\eps_{r,n,k}$ (in both the case $b_t=-1$ and the case $b_t=1$); summing these bounds yields the induction: that our simulator accurately emulates the first $t$ outputs up to statistical distance error $\leq 2t\eps_{r,n,k}$, as desired.
 \end{proof}

\bibliographystyle{plain}
\bibliography{sparsepbib}
\newpage

\end{document}